\documentclass[reprint,amsmath,amssymb,aps,prb,twocolumn,float]{revtex4-2}
\usepackage{graphicx}
\usepackage{graphics}
\usepackage{dcolumn}
\usepackage{bm}
\usepackage{amsfonts}
\usepackage{amssymb}
\usepackage{float}
\usepackage{xcolor}
\usepackage{soul}
\usepackage{natbib}
\usepackage[normalem]{ulem}
\usepackage{soul}

\definecolor{green}{rgb}{0,0.6,0.1}

\newcommand{\RCAST}{Research Center for Advanced Science and Technology, University of Tokyo, 4-6-1 Meguro-ku, Tokyo, 153-8904, Japan}
\newcommand{\Riken}{RIKEN Center for Emergent Matter Science, 2-1 Hirosawa, Wako 351-0198, Japan}
\newcommand{\QPEC}{Quantum Phase Electronics Center, University of Tokyo, Tokyo 113-8656, Japan}
\newcommand{\AP}{Department of Applied Physics, University of Tokyo, Tokyo 113-8656, Japan}
\newcommand{\NTU}{Department of Physics and Center for Theoretical Physics, National Taiwan University, Taipei 10617, Taiwan\looseness=-1}
\newcommand{\NCTS}{Physics Division, National Center for Theoretical Sciences, Taipei 10617, Taiwan\looseness=-1}

\begin{document}

\title{Efficient hydrogen evolution reaction due to topological polarization}

\author{Ming-Chun Jiang$^{1,2}$}      \email{ming-chun.jiang@riken.jp} 
\author{Guang-Yu Guo$^{1,3}$}         

\affiliation{$^1$\NTU}
\affiliation{$^2$\Riken}
\affiliation{$^3$\NCTS}

\author{Motoaki Hirayama$^{2,4}$}          \email{hirayama@ap.t.u-tokyo.ac.jp} 
\affiliation{$^{4}$\QPEC} 
\author{Tonghua Yu$^{5}$}          
\author{Takuya Nomoto$^{6}$}        
\author{Ryotaro Arita$^{2,6}$}        
\affiliation{$^{5}$\AP} \affiliation{$^{6}$\RCAST} 
\date{\today}

\begin{abstract}
Materials carrying topological surface states (TSS) provide a fascinating platform for the hydrogen evolution reaction (HER). Based on systematic first-principles calculations for $A_3 B$ ($A$ = Ni, Pd, Pt; $B$ = Si, Ge, Sn), we propose that topological electric polarization characterized by the Zak phase can be crucial to designing efficient catalysts for the HER. For $A_3 B$, we show that the Zak phase takes a nontrivial value of $\pi$ in the whole (111) projected Brillouin zone, which causes quantized electric polarization charge at the surface. There, depending on the adsorption sites, the hydrogen (H) atom hybridizes with the TSS rather than with the bulk states. When the hybridization has an intermediate character between the covalent and ionic bond, the H states are localized in the energy spectrum, and the change in the Gibbs free energy ($\Delta G$) due to the H adsorption becomes small. Namely, the interaction between the H states and the substrate becomes considerably weak, which is a highly favorable situation for the HER. Notably, we show that $\Delta G$ for Pt$_3$Sn and Pd$_3$Sn are just -0.066 and -0.092 eV, respectively, which are almost half of the value of Pt. 
\end{abstract}

\maketitle

\section{INTRODUCTION} 


Hydrogen evolution reaction (HER) (2H$^+$+2e$^-$ $\rightarrow$ H$_2$) plays a crucial role in renewable energy production for which a variety of experimental and theoretical studies have been performed extensively~\cite{Turner2004,Monai2018}. 
In the celebrated volcano plot (the plot of the efficiency of an electrocatalytic process as a function of the change in the Gibbs free energy $\Delta G$), 
the efficiency takes its maximum at $\Delta G \sim 0$~\cite{Hammer2000,Li2021,Norskov2005}. While it has been shown that the elemental Pt is an ideal catalyst as it is at the top of the volcano plot with $\Delta G$ being merely $\sim$ -0.1 eV~\cite{Norskov2005,Trasatti1972}, Pt has a severe limit for wide-spread industrial usage because of the high cost. Thus 
it is an intriguing challenge to search for a replacement of Pt. Recently, it has been proposed that topological materials having surface states
with high mobility provide a promising platform to study HER~\cite{Xiao2015,Xu2020,Schoop2018,Li2020,Chen2011,Li2019}. 
However, the influence of the topological surface states (TSS) on HER is not fully understood, and strategies to predict new candidates are highly desired.


Weyl semimetals such as 1T'-MoTe$_2$ and the TaAs family were recently studied both experimentally and theoretically as potential HER catalysts \cite{Lu2019,Gupta2017,Rajamathi2017,Ghosh2022}. 
Also, materials with a large Chern number were predicted and turned out to be exceptional catalysts. For example, high turnover frequency of  5.6 and 17.1 H$_2$ s$^{-1}$ were observed for PtAl and PtGa, respectively~\cite{Yang2020}. 
From a TSS perspective, Fermi arcs manifested in Weyl semimetals are characterized by the flow of the Berry curvature between two Weyl points with opposite chirality \cite{Hirayama2018_Topo}.
Meanwhile, 
we can also characterize TSS by the Zak phase, which is defined by integrating the Berry connection along a given reciprocal lattice vector. 
In spinless systems with inversion ($\mathcal{P}$) and time-reversal ($\mathcal{T}$) symmetries, a nontrivial $\pi$ (mod $2\pi$) Zak phase signifies topological surface polarization charge if the system is insulating~\cite{Hirayama2017}. 
In contrast to the surface current originating from Fermi arcs in Weyl semimetals, the $\pi$ Zak phase results in surface dipoles~\cite{Hirayama2017,Hirayama2018_Topo}. 


A distinct advantage of topological materials characterized by the $\pi$ Zak phase is the large size of the relevant region in the projected Brillouin zone (BZ). Even in cases such as PtAl~\cite{Yang2020} where long Fermi arcs connect with each other at the zone boundary, the phase volume of the Fermi arcs is much smaller than that of the region of the $\pi$ Zak phase, which covers a large area in the projected BZ \cite{Hirayama2018_Electrides} and may provide significantly more TSS that are available to assist HER. 
Indeed, it has been recently shown that some nodal line semimetals such as the TiSi family, PtSn$_4$, and VAl$_3$ \cite{Li2017,Li2019a,Kong2021} are promising catalysts. 
However, the effect of the $\pi$ Zak phase on HER is still elusive. One reason is that an interchange of 0 and $\pi$ Zak phase always happens 
in the projected BZ of nodal line semimetals, which makes the topological polarization less significant. 
Therefore, while materials for which the $\pi$ Zak phase lies across the whole projected BZ have yet to be reported, such materials would be an ideal platform to elucidate the role of the topological polarization in HER.


In this paper, we report a nontrivial $\pi$ Zak phase 
over the entire (111) projected BZ for a family of topological semimetals $A_3 B$ (i.e., Pt$_3$Sn, Pd$_3$Sn, Ni$_3$Sn, Ni$_3$Ge, and Ni$_3$Si) \cite{Harris1968,Woo1975,Cannon1984,Suzuki1984,Morozkin2016}, through systematic first-principles calculations. The HER electrocatalytic behavior and its relation to the underlying electronic structure are also investigated. Particularly we focus on the role of TSS, electronegativity, and bonding formation to unveil possible factors for ideal HER electrocatalysts. It should be noted that the HER of $A_3 B$ has not been investigated except for the photocatalytic activity of Pt$_3$Sn (110)~\cite{Yao2019}.

Intriguingly, the $\pi$ Zak phase of $A_3 B$ covers the whole (111) projected BZ due to the absence of double band degeneracy near the Fermi level. Also, this topological phase in $A_3 B$ is stable due to the charge neutrality of each (111) layer. Previous experiments have indeed shown that the (111) termination plane is closely related to the ideal bulk truncation structure \cite{Haner1991,Haner1992}. We further confirm that Pt$_3$Sn and Pd$_3$Sn are exceptional catalysts for HER in that $\Delta G$ is just half of the value of Pt. These results suggest that topological polarization from the $\pi$ Zak phase can serve as a promising selection for HER catalysts. 

The rest of the paper is organized as follows. In Sec. II, a brief description on the crystalline structures of the considered $A_3 B$ compounds in conjunction with the theoretical methods and computational details are given. In Sec. III, we look into the electronic structures, including the topology of the $A_3 B$ family. Also, we show the calculated adsorption energy ($E_{\rm\text{ads}}$), Gibbs free energy ($\Delta G$), and Bader charge analysis to quantify the catalytic behavior for HER. The possible factors affecting the catalytic behavior are discussed by looking into the role of electronegativity, bond formation, and TSS. Finally, conclusions are presented in Sec. IV.    


\section{STRUCTURES and METHODS}

\begin{figure}[tbph] \centering
\includegraphics[width=8.0cm]{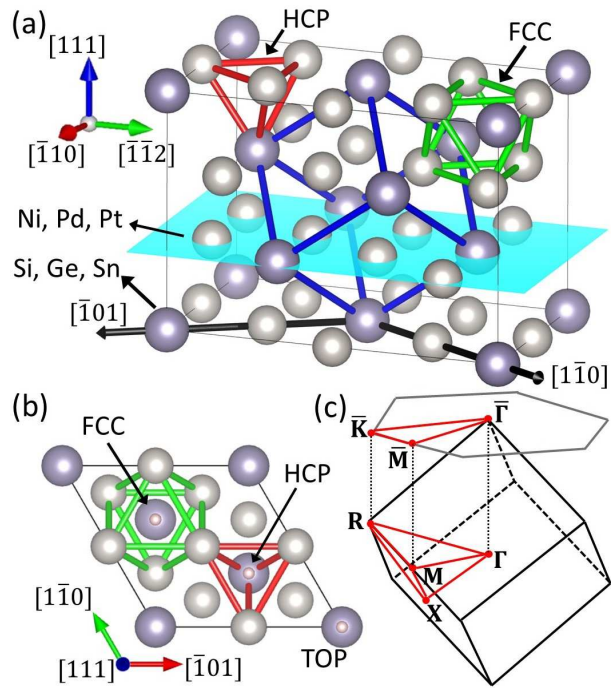}
\caption{Crystal structure, adsorption sites, and Brillouin zone of the $A_3 B$ family. (a) Lattice structure of $A_3 B$ in an orthorhombic supercell. The black vectors show the reduction to a hexagonal supercell. The blue box indicates the primitive cell with the cyan $ab$ facet being the (111) plane of the primitive cell. The red and green bonds form the tetrahedral and octahedral sites, which corresponds to the hcp- and fcc-adsorption site for hydrogen.  
(b) Stable hydrogen adsorption sites on the (111) surface. (c) Brillouin zone of primitive cell and its projection along the [111] direction. Crystal structures are generated via VESTA \cite{Momma2011}.}
\label{fig:struc}
\end{figure}

This study focuses on the five available experimentally synthesized $A_3 B$ compounds, namely $A_3$Sn \cite{Harris1968,Woo1975,Cannon1984}, Ni$_3$Ge \cite{Suzuki1984}, and Ni$_3$Si \cite{Morozkin2016}. All these compounds preserve the cubic symmetry with the space group $Pm\bar{3}m$ (No. 221) with one formula unit per primitive unit cell as illustrated in Fig. \ref{fig:struc}(a). Table \ref{table:struc} lists the experimental lattice constants for all five considered systems. To investigate the surface catalytic behavior of the $A_3 B$ (111) system, we construct a supercell upon the (111) basal planes with cell vectors along [$\bar{1}$01], [1$\bar{1}$0], and [111] 
crystallographic directions, as shown in Fig. \ref{fig:struc}(a)(b); this supercell forms in a hexagonal lattice, and can be readily extended to a slab along the [111] direction [Fig. \ref{fig:struc}(b)]. The BZ of the primitive cell and the (111) projected BZ are depicted in Fig. \ref{fig:struc}(c). 

The $ab$ $initio$ structural optimization, electronic structures, and free energy calculation are based on the density functional theory (DFT) with the generalized gradient approximation (GGA) in the form of Perdew-Burke-Ernzerhof (PBE) \cite{Perdew1996}. The present calculations are performed using the accurate projector-augmented wave method \cite{Blochl1994} implemented in the Vienna $ab$ $initio$ simulation package ({\sc vasp}) \cite{Kresse1993,Kresse1996}. A large plane wave cut-off energy of 400 eV is used. For the BZ integrations with the tetrahedron method \cite{Jepson1993}, Monkhorst-Pack $k$-meshes of 15 $\times$ 15 $\times$ 15 and 9 $\times$ 9 $\times$ 1 are used for the bulk and (111) slab, respectively. The valence orbital set is 6$s^1$5$d^9$ for Pt, 5$s^1$4$d^9$ for Pd, 4$s^1$3$d^9$ for Ni, 4$d^{10}$5$s^2$5$p^2$ for Sn, 3$d^{10}$4$s^2$4$p^2$ for Ge, and 3$s^2$3$p^2$ for Si. We study the nontrivial Zak phase and related topological properties of $A_3 B$ in the absence of spin-orbit coupling (SOC), based on Wannier functions \cite{Wu2018} obtained by \texttt{Wannier90} without the iterative maximal-localization procedure \cite{Marzari1997,Souza2001,Pizzi2020}. We choose the $s$, $p$, and $d$ orbitals for $A$ ($A$= Ni, Pd, Pt) and $s$ and $p$ orbitals for $B$ ($B$= Si, Ge, Sn) as the projectors for Wannier functions. 

\begin{table}[htbp]
\caption{Experimental lattice constants ($a$) of bulk $A_3 B$, as well as the optimized bond length $A$-H and $B$-H for different adsorption sites (fcc, hcp, and top site) on 10-layer (111) $A_3 B$  substrate. The unit is \AA.}
\begin{ruledtabular}
\begin{tabular}{c c c c c c c c}
substrate &     & fcc  &      & hcp  &     & top&      \\
$A_3 B$& $a$ & $A$-H  & $B$-H & $A$-H  & $B$-H & $A$-H & $B$-H \\
\hline
Pt$_3$Sn& 4.004$^a$& 1.83& 3.32& 1.89& 3.39& 3.73& 1.75\\
Pd$_3$Sn& 3.971$^b$& 1.79& 3.31& 1.86& 3.37& 3.58& 1.77\\
Ni$_3$Sn& 3.738$^c$& 1.67& 3.08& 1.73& 3.08& 3.52& 1.77\\
Ni$_3$Ge& 3.571$^d$& 1.67& 2.99& 1.72& 3.05& 3.25& 1.59\\
Ni$_3$Si& 3.500$^e$& 1.66& 2.98& 1.71& 3.05& 3.09& 1.54
\end{tabular}
\end{ruledtabular}
{$^a$Reference \cite{Harris1968}; $^b$Reference \cite{Woo1975}; $^c$Reference \cite{Cannon1984}; $^d$Reference \cite{Suzuki1984}; $^e$Reference \cite{Morozkin2016}}
\label{table:struc}
\end{table}

\begin{table}[htbp]
\caption{Stretching frequency of hydrogen atom on different adsorption sites (fcc, hcp, and top site) of the 10-layer (111) $A_3 B$ substrate. Stretching directions are parallel ($//$) or perpendicular ($\perp$) to the surface. Also, the calculated zero-point energies (ZPE) and the isotropic stretching frequency for hydrogen gas (H$_2$) are listed. For comparison, we also list the results of hydrogen adsorbed Pt (111) systems and related studies. Units are meV.}
\begin{ruledtabular}
\begin{tabular}{c c c c c c c c c c}
substrate& fcc  &      &    &hcp   &        &    &  top&        &    \\
$A_3 B$& $//$ & $\perp$& ZPE& $//$ & $\perp$& ZPE& $//$& $\perp$& ZPE \\
\hline
Pt$_3$Sn& 126& 121& 187& 85& 136& 153& 49& 211& 155\\
Pd$_3$Sn& 136& 115& 194& 103& 127& 167& 40& 201& 140\\
Ni$_3$Sn& 144& 119& 204& 119& 128& 183& 36& 202& 137\\
Ni$_3$Ge& 144& 133& 211& 122& 146& 195& 43& 222& 155\\
Ni$_3$Si& 142& 138& 211& 119& 152& 195& 40& 233& 156\\
Pt      &  73& 150& 148&    &    &    & 50& 281& 190 \\
        &    &    & 134$^a$ &    &    &   & & & 182$^a$ \\
\hline
 & Isotropic& ZPE& & & & & & & \\
\hline
H$_2$ & 515& 258& & & & & & & 
\end{tabular}
\end{ruledtabular}
{$^a$Reference \cite{Hanh2014} (DFT)} 
\label{table:ZPE}
\end{table}

The $A_3 B$ (111) slab is modeled with 10 atomic layers and a 14 {\AA} vacuum layer. We use the $1\times1$ supercell to study the adsorption, corresponding to 1 monolayer coverage of H. Adsorption occurs on one side of the slab, and the top 4 metal layers are allowed to relax along with the adsorbate in these calculations. The resulting stable adsorption sites of $A_3 B$ (111) are shown in Fig. \ref{fig:struc} (b), which are the $B$ top site (top-adsorption), the $A_3$ tetrahedral site (hcp-adsorption), as well as the $A_3$ octahedral site (fcc-adsorption). The optimized bond length between the adsorbed H and the substrate on different adsorption sites is listed in Table \ref{table:struc}. 

The adsorption energy ($E_{\rm\text{ads}}$) and the difference in the Gibbs free energy $\Delta G$ for H adsorption on $A_3 B$ (111) are used to evaluate the catalytic behavior toward HER \cite{Norskov2005,Rajamathi2017,Kong2021}. The equations go by \cite{Norskov2005}: 
\begin{equation} \label{eq_Ea}
    E_{\rm\text{ads}} = \frac{1}{n}(E_{\rm\text{slab+H}}-E_{\rm\text{slab}}-\frac{n}{2} E_{\rm\text{H}_2})
\end{equation}
where $n$ is the number of H atoms, $E_{\rm\text{slab+H}}$, $E_{\rm\text{slab}}$, and $E_{\rm\text{H}_2}$ are the total energies of the H atoms adsorbed on the slab, clean slab, and gas-phase H$_2$, respectively; also,  
\begin{equation} 
\Delta G = \Delta E_{\rm\text{ads}} + \Delta E_{\rm\text{ZPE}} - T\Delta S 
\end{equation}
where $E_{\rm\text{ads}}$ is Eq. \ref{eq_Ea} , $\Delta E_{\rm\text{ZPE}}$ and $\Delta S$ are the zero-point energy difference and the entropy difference between the adsorbed H states and the gas-phase H$_2$, respectively. $\Delta E_{\rm\text{ZPE}}$ is calculated by the difference in vibrational frequency ($\omega$) of H and gas-phase H$_2$ via $E_{\rm\text{ZPE}}=\sum_i \frac{1}{2}\hbar\omega$. For the entropy difference, due to the small entropy difference between the two phases, we could approximate it by \cite{Norskov2005}  
\begin{equation}
    \Delta S \approx -\frac{1}{2}S^0_{\rm\text{H}_2}, \quad S^0_{\rm\text{H}_2} = 130.674 \quad \rm\text{J/Kmol}
\end{equation}
where S$^0_{\rm\text{H}_2}$ is the standard entropy value of gaseous H$_2$ at 300K \cite{Chase1998}. Therefore, the $T \Delta S$ is roughly 0.2 eV.    

\section{RESULTS AND DISCUSSION}
\subsection{Electronic Structures of $A_3 B$}

\begin{figure*}[tbph] \centering
\includegraphics[width=\textwidth]{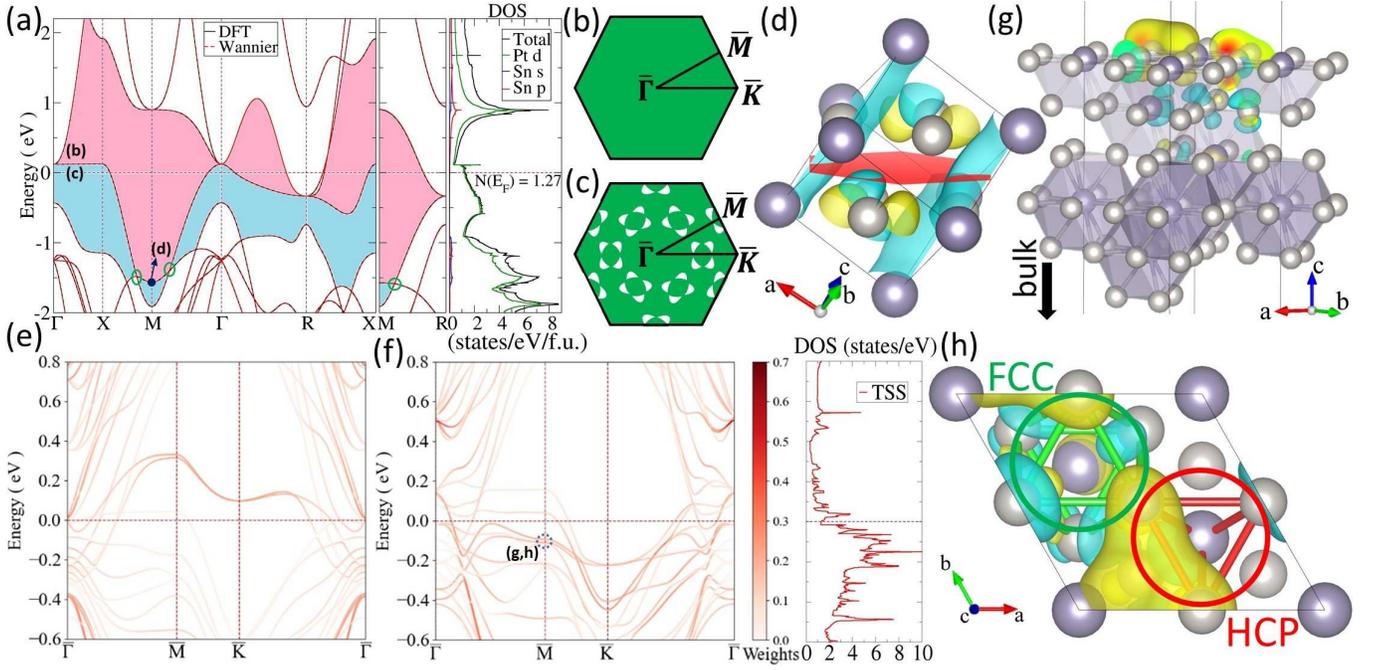}
\caption{Electronic structure of Pt$_3$Sn bulk and (111) slab in the absence of the spin-orbit coupling. (a) Electronic band structure and density of states of bulk Pt$_3$Sn, with the high-symmetry $\textbf{k}$ points shown in Fig. \ref{fig:struc}(c). The calculated Zak phase on the (111) projected Brillouin zone of the (b) pink and (c) blue gap. The green region indicates the $\pi$ Zak phase. (d) Real bulk wave function of the eigenstate at the time-reversal invariant $M$ point pointed in black in (a). Yellow and blue represent the positive and negative components, respectively. The red slice in the middle indicates the (111) plane. (e)(f) Projected electronic band structure and density of states for the (111) surface of a 10-layer slab of Pt$_3$Sn in the (e) tight-binding Hamiltonian and (f) the GGA. The color bar indicates the weights of the topological surface state. The Fermi level is set to 0 regarding all band structures. (g)(h) Real wave functions of topological surface state at the $\bar{M}$ point as circled in blue in (f). Yellow and blue represent the positive and negative components, respectively.}
\label{fig:elec}
\end{figure*}

First of all, we look into the electronic structures of $A_3 B$. As stated in the previous section, $A_3 B$ crystallizes in centrosymmetric simple cubic structure with the space group $Pm\bar{3}m$ (No. 221). Due to the similarity in electron configurations, 
the same topological properties can be seen for all $A_3 B$ materials. Also, the electronegativity of the nickel family and carbon family are close (Pt = 2.28, Pd = 2.2, Ni = 1.91; Sn = 1.96, Ge = 2.01, Si = 1.9 \cite{Zumdahl2004}), indicating that covalent bonds may form in $A_3 B$. On the surface, covalent bonds are cut and leave dangling bonds, giving rise to possible surface states. Here, we look into the details of the electronic structure of Pt$_3$Sn as a representative compound of $A_3 B$.     

We calculate the electronic band structure of Pt$_3$Sn in the absence of SOC, and the results are presented in Fig. \ref{fig:elec}. First of all, Fig. \ref{fig:elec}(a) shows the electronic band structure as well as the DOS near the Fermi level of bulk Pt$_3$Sn. The low DOS and few crossings at the Fermi level suggest the semi-metallic nature. 
Bands around the Fermi level are contributed by the Pt $d$ orbitals with minor contributions of the Sn $s$ and $p$ orbitals. We, therefore, construct the Wannier functions utilizing the mentioned orbitals with the interpolated band structure plotted in dashed red lines in Fig. \ref{fig:elec}(a). We note that the band structures from DFT and Wannier functions are identical, both of which agree well with previous work in the GGA \cite{Kim2017}. 
Also, due to the $\mathcal{PT}$ symmetry, all bands are Kramers degenerate. Henceforth, we discuss the band degeneracy for one spin channel. 

From Fig. \ref{fig:elec}(a), we notice that on the $\Gamma$-$R$ $C_{3v}$-symmetric line of Pt$_3$Sn, a nondegenerate band crosses a doubly degenerate band, 
forming two spinless triple nodal points on $\Gamma$ and $R$. The same behavior of splitting arises along the $C_{4v}$-symmetric lines $\Gamma$-$X$ and $R$-$M$ \cite{Ahn2018}. In a $\mathcal{PT}$-symmetric system, doubly degenerate points tend to form in loops, or the so-called nodal lines, characterized by the $\pi$ Berry phase mechanism and related to the Zak phase mechanism \cite{Hirayama2018_Topo}. Interestingly, in the present case, triple nodal points appear and do not form into loops. The split-nodal lines on the zone boundary are also mainly unoccupied. Consequently, we see significant gaps in the BZ outside of the triple nodal points around the Fermi energy (blue and pink gap in Fig. \ref{fig:elec}(a)). This leads us to expect the pink gap to have identical Zak phase across the whole projected BZ. Our requirement for an ideal platform to study the relation between TSS and the surface reaction is then fulfilled.

Aside from the triple nodal point, in Pt$_3$Sn, we observe doubly degenerate points along $M$-$\Gamma$, $M$-$X$, and $M$-$R$ line at -1.4 $\sim$ -1.5 eV (circled in green in Fig. \ref{fig:elec}(a)). In the BZ, this forms into two loops protected by mirror symmetry of the (110) and (001) planes. As discussed above, we would now expect the Zak phase property originating from the nodal ring around the $M$ point in the blue gap to span across the pink gap and, more importantly, the majority of the Fermi level. To confirm this idea, using the constructed Wannier functions, we calculate the Zak phase by integrating the Berry connection along a reciprocal lattice vector \textbf{G}. In the case of interest, it is the [111] direction that indicates the surface polarization charge on the (111) plane. The choice of the unit cell will be given in Fig. S1 of the Supplemental Material (SM)\cite{SM}. We then calculate the Zak phase as follows, 
\begin{equation}
    \theta(\textbf{k}_\parallel)=-i\sum^{\rm\text{occ.}}_n\int^{|\textbf{G}|}_0 d\textbf{k}_\perp \left<u_n(\textbf{k})|\nabla_{\textbf{k}_\perp}|u_n(\textbf{k})\right>,
\end{equation}
where $u_n(\textbf{k})$ is the periodic part of the bulk Bloch wave function. The wave vector \textbf{k} is decomposed into the component along the lattice vector \textbf{G} (i.e., $\textbf{k}_\perp$) and the one perpendicular to $\textbf{G}$ (i.e., $\textbf{k}_\parallel$). 

The calculated results are given in Fig. \ref{fig:elec} (b) for the pink gap and (c) for the blue gap in Fig. \ref{fig:elec}(a). In Fig. \ref{fig:elec}(b), the Zak phase along the [111] direction is equal to $\pi$ in the entire projected BZ. This suggests surface polarization charges equal to $e$/2 per surface unit cell on the (111) surface \cite{Hirayama2018_Electrides}. On the other hand, in Fig. \ref{fig:elec}(c), we notice patches of 0 Zak phase appearing around 12 centers with, i.e., six on $\bar{M}$. These 12 centers correspond to the 12 $M$ points in the cubic BZ (Fig. \ref{fig:struc}(c)), and the butterfly-shaped patches are the projections of two nodal rings formed around $M$ ( green circles in Fig. \ref{fig:elec}(a)). We note that full Zak phase coverage has not been shown in semimetals, although it has been reported in few insulators such as Sc$_2$C \cite{Hirayama2018_Electrides}. Therefore, the $A_3 B$ family or, in general, the $Pm\bar{3}m$ space group \cite{Ahn2018} could be a promising platform to study Zak phase physics in semimetals.

While the Zak phase is a bulk property, we can illustrate the origin of surface polarization charge by the wave function of the bulk eigenstates. Fig.~\ref{fig:elec}(d) shows the wave function of the highest valence band at the $M$ point, which is in the gap of interest. We see the Pt 5$d$ and Sn $p$-like orbitals extending across the (111) plane in the middle. This indicates that the surface charges originate from dangling bonds as (111) termination plane is formed. 
Furthermore, the protected surface charges would cause deficiency of the bulk charges. Due to charge conservation, such loss will be compensated by the emerging surface band in a gap for a thick (111) Pt$_3$Sn slab, which is shown in the slab band structure of the constructed Wannier functions (Fig.~\ref{fig:elec} (e)).
While the degenerate points occur in $\Gamma$ and $R$ point of the bulk band structure Fig.~\ref{fig:elec} (a), the surface band at their projection, the $\bar{\Gamma}$ point, would merge with the bulk band. However, the large gap at the $X$ and $M$ point would allow us to observe the highlighted TSS appearing on the high symmetry $\bar{M}$ and $\bar{K}$ point.

\begin{table*}[t]
\caption{The adsorption energy ($E_{\rm\text{ads}}$), Gibbs free energy ($\Delta G$), and Bader charge analysis of H on different adsorption sites (fcc, hcp, and top site) of the 10-layer (111) $A_3 B$ substrate in the absence of the spin-orbit coupling at room temperature. Also, the Bader charge analysis results of the top layer atoms are listed with charge differences before and after the adsorption given in the row below. Note that the number of charges for $A$ is the average between the three atoms on the surface.}
\begin{ruledtabular}
\begin{tabular}{c | c c |c c c| c c |c c c |c c |c c c }
substrate&  fcc& (eV)  &  &   & ($e^-$)  &  hcp& (eV)  &    &    & ($e^-$)  & top& (eV)  &    &    & ($e^-$)   \\
$A_3 B$   &  $E_{\rm\text{ads}}$& $\Delta G$ &  H & $A$ & $B$ & $E_{\rm\text{ads}}$& $\Delta G$ &  H  & $A$ & $B$ & $E_{\rm\text{ads}}$& $\Delta G$ &  H  &  $A$ & $B$ \\
\hline
Pt$_3$Sn& -0.324& -0.066& 1.059& 10.342& 13.027& -0.606& -0.382& 1.067& 10.338& 13.018& 0.828& 1.055& 1.276& 10.377& 12.962\\
         & -0.44$^a$  &   & +0.059 &  -0.048 & -0.026  &  &   &  +0.067  &  -0.051  & -0.035  & &   & +0.276 & -0.012& -0.091\\   
Pd$_3$Sn& -0.356& -0.092& 1.099& 10.212& 13.293& -0.668& -0.430& 1.142& 10.218& 13.263& 0.999& 1.211& 1.297& 10.250& 13.197\\
        &       &       & +0.099& -0.043& -0.021&  &   & +0.142& -0.037& -0.052& &   &  +0.297& -0.005& -0.118\\
Ni$_3$Sn& -0.638& -0.363& 1.244& 10.070& 13.602& -0.915& -0.661& 1.270& 10.071& 13.585& 0.927& 1.136& 1.290& 10.153& 13.446\\
        &  &   & +0.244& -0.076& -0.054&  &   & +0.270& -0.074& -0.070& &   & +0.290& +0.007& -0.209\\
Ni$_3$Ge& -0.544& -0.262& 1.225& 9.991& 13.838& -0.978& -0.712& 1.262& 9.986& 13.835& 0.823& 1.049& 1.282& 10.067& 13.684\\
        &  &   & +0.225& -0.071& -0.056&  &   & +0.262& -0.076& -0.060& &   & +0.282& 0.005& -0.210\\
Ni$_3$Si& -0.496& -0.214& 1.220& 10.022& 3.738& -0.991& -0.725& 1.249& 10.024& 3.712& 0.630& 0.857& 1.510& 10.115& 3.360\\
        &  &   & +0.220& -0.076& -0.044&  &   & +0.249& -0.075& -0.070& & & +0.510& +0.016& -0.422\\
\hline
Pt & -0.362& -0.143& & & & & & & & & -0.391& -0.130 \\
  & -0.39$^b$& & & & & & & & & & -0.42$^b$& 
\end{tabular}
\end{ruledtabular}
{$^a$Reference \cite{Liu2003} (DFT) ; $^b$Reference \cite{Hanh2014} (DFT)} 
\label{table:Gibbs}
\end{table*}

With the confirmation of the topological nature of Pt$_3$Sn, we perform the self-consistent DFT calculation of the slab band structure of a 10-layer (111) Pt$_3$Sn slab. The result is depicted in Fig. \ref{fig:elec} (f). By comparing Figs. \ref{fig:elec}(e) and (f), we can see that the TSS merge into the bulk states at an energy level around $-0.1 \sim -0.2$ eV as predicted by the DFT, as clearly shown in the left panel of Fig. \ref{fig:elec}(f). This is due to the impact of surface charge redistribution, which is adequately treated in DFT but ignored in the calculation based on the Wannier functions. Nonetheless, such TSS still exist in the vicinity of the Fermi level and would contribute to the surface reaction (see the projected density of states shown in the right panel of Fig.~\ref{fig:elec}(f)). Furthermore, we plot out the real part of the wave function at the time-reversal invariant point $\bar{M}$ of 10-layer Pt$_3$Sn (111) at Figs. \ref{fig:elec}(g) and (h). We notice that the TSS are contributed mainly by the Pt orbitals with hybridization between the three Pt atoms on the surface. Moreover, the TSS cover the hcp site partially but is repelled from the fcc site, as indicated by the red and green circles, respectively in Fig. \ref{fig:elec}(h).

Note that the electronic structures of other $A_3 B$ compounds are in Fig. S2-S5 of the SM. The difference in the electronic structure is due to the difference in electronegativity, especially as site $A$ changes. Nevertheless, the $\pi$ Zak phase on the (111) projected BZ is seen among all $A_3 B$. The resulting TSS are also in the vicinity of the Fermi level (Fig. S2-S5(e)).
It is also essential to point out that the inclusion of SOC would cause a topological phase transition for $A_3 B$ to go from a $\pi$ Zak phase semimetal into a weak Dirac dispersion along [111] direction \cite{Kim2017}. Bulk and surface electronic structures in the spinful case are provided in Fig. S6 of the SM. With these, in the Appendix, we discuss the effect of SOC where we show that SOC hardly affects the interaction between the surface states and the H adsorption. 

Now that we have shown the topological polarization in Pt$_3$Sn, we will proceed to the catalytic behavior. The role and the effect of the TSS coverage among different adsorption sites will be further clarified in Sec. III C and D, respectively.

\subsection{Catalytic behavior of $A_3 B$}

To investigate the catalytic behavior of $A_3 B$, we calculate the $E_{\rm\text{ads}}$ as well as the $\Delta G$ of H residing at the fcc, hcp, and top site in the absence of SOC. Results are organized in Table \ref{table:Gibbs}. First of all, intriguingly, the $\Delta G$ of Pt$_3$Sn and Pd$_3$Sn (111) fcc-adsorption is of value $-0.066$ eV and $-0.092$ eV, respectively. With $\Delta G$ roughly half the value of Pt ($-0.143$ eV), Pt$_3$Sn and Pd$_3$Sn are thus prominent HER catalysts with exceptional low $\Delta G$ and less or even zero composition of Pt. Secondly, Ni$_3 B$ (111) have $\Delta G$ of value around $-0.25$ eV on the fcc site, larger than that of Pt$_3$Sn and Pd$_3$Sn, indicating a worse catalytic behavior. Nevertheless, they are comparable with related theoretical calculations in GGA such as nodal line semimetals PtSn$_4$ ($-0.28$ eV) \cite{Li2019a}, VAl$_3$ (0.25 eV) \cite{Kong2021}, or Weyl semimetal TaP ($-0.40$ eV) \cite{Rajamathi2017}. By results of $\Delta G$, we predict the considered $A_3 B$ family (Pt$_3$Sn, Pd$_3$Sn, Ni$_3$Sn, Ni$_3$Ge, Ni$_3$Si) could be promising catalysts for HER with much less Pt consumption. The calculated results (Table \ref{table:Gibbs}) agree well with previous studies \cite{Liu2003,Hanh2014}.

Among different adsorption sites, the fcc site has the lowest $\Delta G$ with a value around $-0.1$ $\sim$ $-0.3$ eV. Adsorptions on the hcp site have $\Delta G$ of value about $-0.4$ $\sim$ $-0.7$ eV, and the top site behaves the worst in catalysis with $\Delta G$ of nearly 1 eV. This indicates that the H atoms are more prone to the fcc-site adsorption while being easier to detach simultaneously, which leads to better-performance catalytic behavior. Note that for the fcc and hcp sites, Pt$_3$Sn and Pd$_3$Sn have lower $\Delta G$ than that of Ni$_3 B$. However, on the top site, the $\Delta G$ of Ni$_3$Si is surprisingly the lowest. This could be attributed to the absence of outer $d$ orbitals in the case of top-adsorption, as supported by our results that $p$ orbitals could be better for active sites with only one bond. Still, $\Delta G$ of the top site is at least 2 to 4 times larger than those of the fcc and hcp sites, which makes it less relevant in the catalytic process. The feature that distinguishes the fcc and hcp sites from the top site can be seen in Fig. \ref{fig:elec} (h). The TSS protected by the nontrivial Zak phase have coverage mostly around the fcc and on the hcp site, suggesting a possible tendency of adsorption sites with TSS having superior catalytic efficiency.

\begin{figure}[b] \centering
\includegraphics[width=8.6cm]{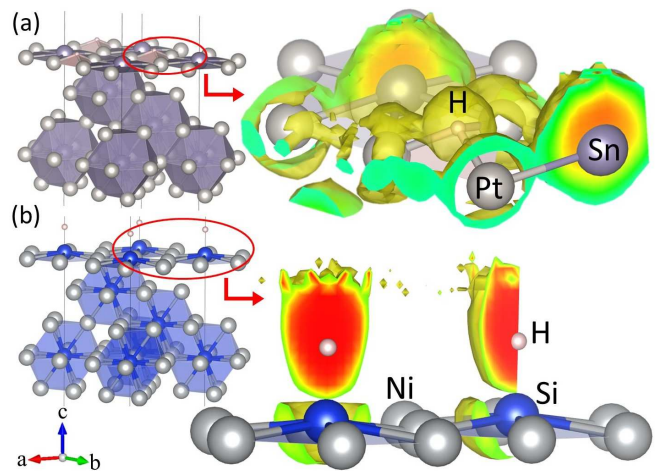}
\caption{Electron localized functions for (a) Pt$_3$Sn (111) fcc-adsorption and (b) Ni$_3$Si (111) top-adsorption. }
\label{fig:elf}
\end{figure}

\begin{figure*}[tbph] \centering
\includegraphics[width=\textwidth]{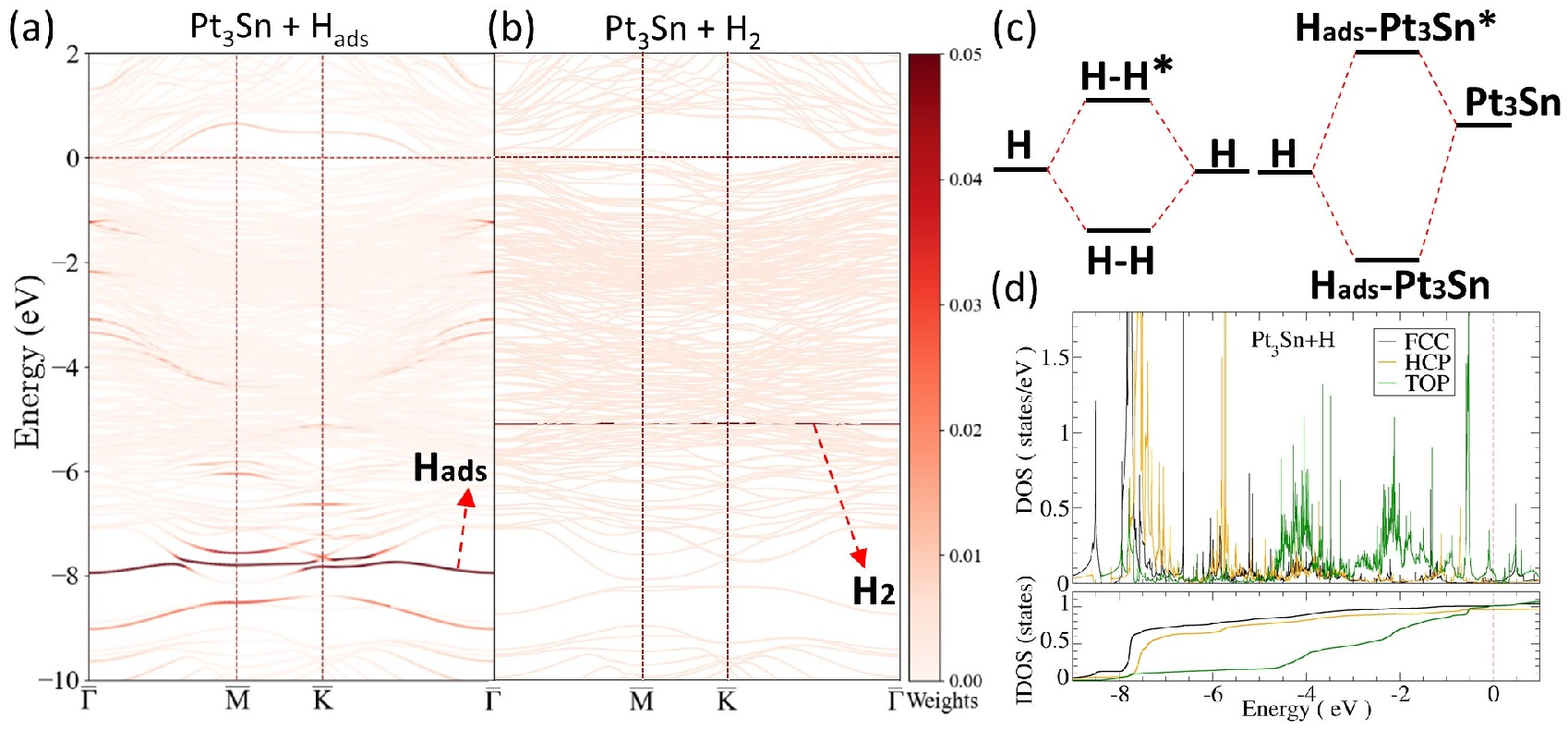}
\caption{Bonding between Pt$_3$Sn (111) and H atom. (a) Band structure of 10-layer Pt$_3$Sn (111) fcc-adsorption highlighted with states projected onto the H atom. (b) Band structure of 10-layer Pt$_3$Sn (111) with a H$_2$ molecule sufficiently away from the substrate. The energy level of the H$_2$ molecule is highlighted in brown. (c) Schematic of covalent bonding and an intermediate between the covalent and ionic bonding. (d) Density of states of the H atom for 10-layer Pt$_3$Sn (111) fcc-, hcp-, and top-adsorption. The cumulative sum of the density of states is shown in the lower panel. }
\label{fig:bond}
\end{figure*}

Finally, we also perform the Bader charge analysis \cite{Tang2009} to investigate the charge transfer behavior, and the results are organized in Table \ref{table:Gibbs}. Each column beside $\Delta G$ shows the number of electrons that belong to the top layer atoms. The charge difference before and after the adsorption is given in the next row. Firstly, we notice that the H atom tends to receive electrons from $A_3 B$ during the adsorption. Secondly, the H atom that receives fewer electrons shows a propensity for a lower $\Delta G$. For example, the H atom for Pt$_3$Sn and Pd$_3$Sn fcc-adsorption, with the lowest $\Delta G$, receives only 0.059 $e^-$ and 0.099 $e^-$, respectively. On the other hand, the H atom of Ni$_3 B$ fcc-adsorption receives $\sim$ 0.25 $e^-$. 
In general, the H atom on the fcc site would receive fewer electrons among the same material than that on the top site. Moreover, among the same adsorption sites, the H atom on the Pt$_3$Sn and Pd$_3$Sn would also receive fewer electrons than the Ni$_3 B$. From the observations above, we may propose that the conditions of high catalytic performance for the $A_3 B$ family are that the H atom interacts with the TSS significantly and receives a small number of electrons from the substrate. We would further incorporate the impact of electronegativity, bonding type, and the TSS in the following subsections based on such an argument.

\subsection{Role of electronegativity and bonding type}

The number of electrons transferred during the chemical reaction can be estimated by electronegativity and further classified into different bonding types such as the covalent or the ionic bonding. 
From the viewpoint of electronegativity, we notice that for $A_3 B$ fcc- and hcp-adsorption, the closer the electronegativity for $A$ is to that of H atom (Pt = 2.28, Pd = 2.2, Ni = 1.91, H = 2.2 \cite{Zumdahl2004}), the less electrons are donated and the smaller the $\Delta G$ gets, which indicates the better the catalytic behavior is. The underlying mechanism is the shifting of the $d$-band center as interpreted by the ratio of electronegativity between $A$ and H \cite{Xu2018,Gao2020,Li2021}. 



A covalent bond is formed if the Bader charge of H remains as 1 $e^-$ after the adsorption. Table \ref{table:Gibbs} shows that, after adsorption, the electrons received by the H atom range from 1.05 $e^-$ to 1.5 $e^-$, indicating an intermediate bonding between the covalent and ionic. The Sabatier principle \cite{Li2021,Sabatier1920} claims that during an electrocatalytic process, the bonding should neither be too strong nor too weak to avoid difficulty for the reactant to either attach to or detach from the catalysts. Therefore, to understand the bonding character \cite{Becke1990,Silvi1994}, we calculate the electron localized function (ELF) for the two cases with the least [Fig. \ref{fig:elf}(a), Pt$_3$Sn fcc-adsorption, 1.059 $e^-$] and most [Fig. \ref{fig:elf}(b), Ni$_3$Si top-adsorption, 1.510 $e^-$] electrons that the H atom obtains. In Fig. \ref{fig:elf}(a), the ELF appears in the vicinity of the Pt, Sn, and H atoms, suggesting the bonding nature is more of the covalent type. On the other hand, in Fig. \ref{fig:elf}(b), the electrons are more accumulated around the Si and H atom itself with a hole along the bonding direction. This suggests the bonding nature is more of ionic bonding.     

We have shown that the $A_3 B$ family forms an intermediate between covalent and ionic bonding after the adsorption with the degree determined by the electronegativity. While it seems like the more covalent the bonding is, the better the catalytic behavior is for the fcc- and hcp-site, the connection to the Sabatier principle is not clear. Therefore, we attempt to elucidate such connections from the band theory point of view. If an atom is weakly attached to a substrate, the overall band structures can be simply viewed as a linear combination of both individual bands. 

In Fig. \ref{fig:bond}(a), the projected band structures of the H atom of the Pt$_3$Sn (111) fcc-adsorption are depicted. First, we notice that the H state is highly localized at an energy level of $\sim$ $-8$ eV. Such high localization indicates a weak interaction between Pt$_3$Sn and the H atom. Also, the bonded H state deeply buried in the spectrum indicates a low $d$-band center, implying the weakening of adsorption energy by the filling of the anti-bonding state \cite{Zheng2014}. Secondly, we further demonstrate the intermediate bonding formation by calculating the band structure of the Pt$_3$Sn slab with an isolated H$_2$ molecule, and the results are in Fig. \ref{fig:bond}(b). The brown line at $\sim$ $-5$ eV denotes the energy level of the H$_2$ molecule, which is equivalent to a fully covalent bonding situation. 
If the energy level of H in the case of Pt$_3$Sn+H is lower than the case of Pt$_3$Sn+H$_2$, the bonding of Pt$_3$Sn+H would be closer to a covalent bonding. 
Indeed, the highlighted states in Figures \ref{fig:bond}(a) and (b) show a large energy drop, suggesting that the bonding type of Pt$_3$Sn+H is closer to a covalent pattern. Figure \ref{fig:bond}(c) is a schematic of the covalent bonding of the H$_2$ molecule and the intermediate bonding between Pt$_3$Sn and the H atom. The energy level difference between Pt$_3$Sn and H also explains the difference in the weighting of H between the bonding and anti-bonding state, which is why we observe a strong bonding state of H in Fig. \ref{fig:bond}(a) in $\sim$ $-8$ eV but a less clear anti-bonding state in the higher energy spectrum in $\sim$ 0.6 eV. 

Furthermore, we illustrate the connection between the localization of the H bands and the catalytic behavior by investigating the DOS. Figure \ref{fig:bond}(d) shows the DOS of the H atom in the cases of Pt$_3$Sn (111) fcc-, hcp-, and top-adsorption. From the top panel of Fig. \ref{fig:bond}(d), we notice the black peak at $\sim$ $-8$ eV for the fcc-adsorption, which is the dark band in Fig. \ref{fig:bond}(a). Interestingly, unlike fcc-adsorption, where no other peaks are shown across the energy spectrum, we see a second prominent peak around $-6$ eV for hcp-adsorption and apparent dispersive bands across the energy spectrum for top-adsorption. From the lower panel of Fig. \ref{fig:bond}(d), the cumulative sum of the DOS is presented. The sharp slope of the black curve indicates the highly localized H bands, while the green curve represents the highly hybridized H bands. Figures S7 are results for other $A_3 B$, and the slope feature is identical to that of Pt$_3$Sn. From Table \ref{table:Gibbs}, the $\Delta G$ of the fcc-adsorption is the lowest, with the top-adsorption being the highest. Therefore, the Sabatier principle is interpreted in the band theory view of point by arguing that the highly localized bands indicate a merely interacting H atom and the slab. This would then lead to better catalytic behavior, as shown from the $\Delta G$ calculation.

\subsection{Role of topological surface states}

\begin{figure*}[tbph] \centering
\includegraphics[width=\textwidth]{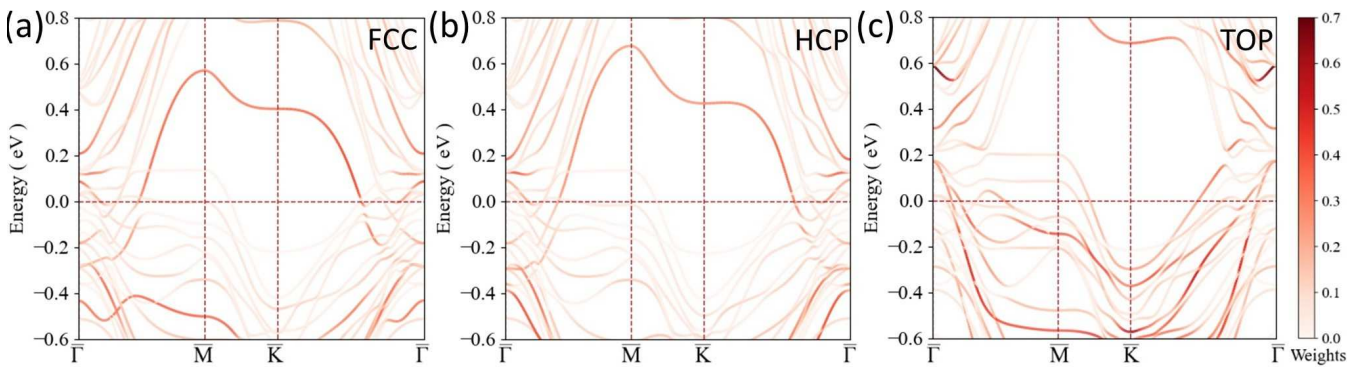}
\caption{Surface electronic structures of Pt$_3$Sn (111) with H adsorption onto the (a) fcc site, (b) hcp site, and (c) top site.}
\label{fig:projected}
\end{figure*}

Now we would like to turn our attention back to the role of TSS in the $A_3 B$ system. From Table \ref{table:Gibbs} and Fig. \ref{fig:elec}(h), we have shown more TSS coverage at the fcc and hcp sites than at the top site, and accordingly, the $\Delta G$ of the top site is the largest. In Fig. \ref{fig:projected}, we calculate the slab band structure of the 10-layer Pt$_3$Sn (111) fcc-, hcp-, and top-adsorption. The highlighted states are the TSS, and thus by comparing Fig. \ref{fig:elec}(f) and Fig. \ref{fig:projected} we can understand the difference between TSS before and after the adsorption. Also, results of other $A_3 B$ are in Figs. S8-S11.  

\begin{figure*}[tbph] \centering
\includegraphics[width=0.8\textwidth]{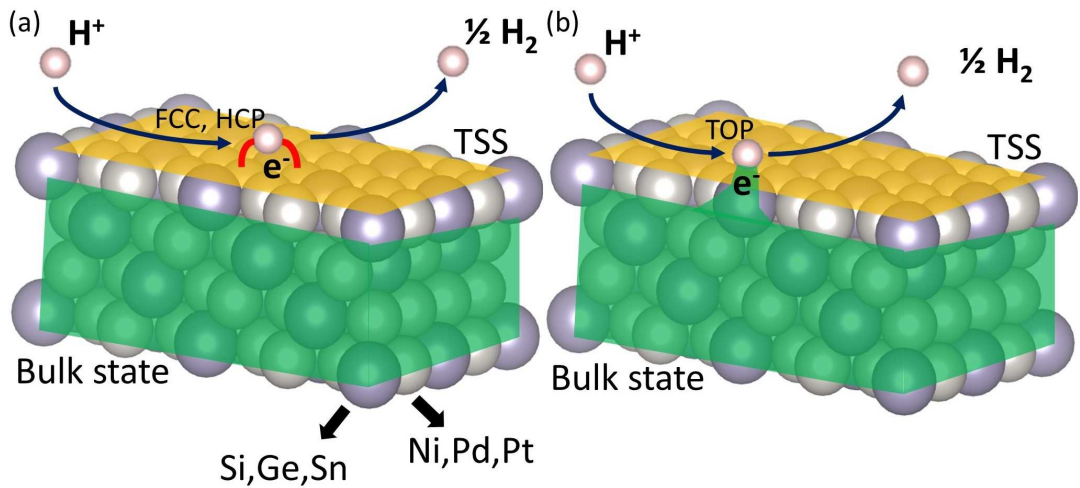}
\caption{Schematic plot of interaction between H atom with (a) the topological surface states and (b) the bulk states of the $A_3 B$ family.}
\label{fig:schematic}
\end{figure*}

Interestingly, the TSS shift upward from $\sim$ $-0.2$ eV to an energy level above the Fermi energy and disperses across 0 to 0.6 eV and 0.8 eV for the fcc- and hcp-adsorption, respectively [Figs. \ref{fig:projected}(a) and (b)]. The up-shift of TSS implies the donation of electrons from the Pt$_3$Sn substrate to the H atom, and more importantly, the electrons are mainly donated from the TSS. This implication coincides with our Bader charge calculation. For example, from Table \ref{table:Gibbs}, in the case of Pd$_3$Sn fcc-adsorption, H receives a total amount of 0.099 electrons, which is roughly coming from the top layer Pd atoms ($-0.129$ $e^-$). Even in cases with larger $\Delta G$, for example, Ni$_3$Sn hcp-adsorption, H receives 0.270 electrons, which can be covered by the top layer Ni atoms ($-0.222$ $e^-$). However, we notice less significant up-shifts of the TSS from Fig. \ref{fig:projected}(c) for the top-adsorption. Although the states around $-0.2$ eV are more localized than before the adsorption, the TSS are less affected than in fcc- or hcp-adsorption. From the Bader charge analysis (Table \ref{table:Gibbs}), the number of electrons that H receives during the Pt$_3$Sn top-adsorption is 0.276, which does not fully originate from the top surface ($-0.127$ $e^-$ in total). This trend appears across all top-adsorption of the considered $A_3 B$. Importantly, this indicates that H interacts with the bulk bands on the sites that underperform the catalytic process. The interaction between the H atom and the bulk bands on the top site also explains the more dispersive and, therefore, more hybridized H DOS in Fig. \ref{fig:bond}(d). 

The role of the TSS on HER proposed in this study is illustrated in Fig. \ref{fig:schematic}. From Fig. \ref{fig:schematic}(a), the role of TSS is demonstrated via the case of the fcc- and hcp-adsorption in the $A_3 B$ family. The $\pi$ Zak phase in the $A_3 B$ covers the whole (111) projected BZ, which guarantees the emergence of the (111) surface dipole. The resulting TSS cover primarily the fcc- and hcp-site (Fig. \ref{fig:elec}(h)). When the H atom interacts with the TSS (Fig. \ref{fig:projected}(a)(b)), the resulting catalytic behavior is better (Table \ref{table:Gibbs}) due to the weakly hybridized H bands [Fig. \ref{fig:bond}(a)(d)] and accordingly, a weak intermediate between the covalent and ionic bonding is formed. On the other hand, in the case of top-adsorption, as illustrated in Fig. \ref{fig:schematic}(b), the catalytic behavior is significantly worsened when the H atom interacts with the bulk bands \ref{fig:projected}(c). The resulting H bands disperse across the energy spectrum as presented in Fig. \ref{fig:bond}, indicating a stronger hybridization between the substrate and the H atom. This leads to stronger bonding, leading to higher reaction energy needed for the HER. The proposed mechanism suggests that flat H bands are an indicator of suitable catalysts while TSS are good weakly interacting sites. We expect materials with topological polarization characterized by the $\pi$ Zak phase to be a fruitful platform for future catalysis and surface science studies.          

\section{CONCLUSIONS}
To summarize, we have investigated the HER behavior and its relation to the underlying 
electronic structure of topological semimetals $A_3 B$ ($A$= Ni, Pd, Pt; $B$= Si, Ge, Sn), based on first-principles calculations. 
First of all, we find that the spinless $A_3 B$ family has a nontrivial $\pi$ Zak phase across the entire (111) projected BZ. The Zak phase originates from a nodal ring around the $M$ point at $-1.5$ eV, and the lack of doubly degenerate points allows the $\pi$ Zak phase to span across the whole projected BZ.   
The TSS appear around $\sim$ $-0.2$ eV in the slab band structure with real space surface charges covering partially the hcp site and the rim of the fcc site. 
Excitingly, the $\Delta G$ calculation predicts $A_3 B$ to be exceptional catalysts for HER with very small $\Delta G$ value of $-0.066$ eV and $-0.092$ eV for Pt$_3$Sn and Pd$_3$Sn fcc-adsorption, respectively. 
This is roughly half the value of the ideal catalyst Pt fcc-adsorption ($-0.143$ eV). 
The rest, i.e., Ni$_3 B$, has $\Delta G$ values around $-0.2$ $\sim$ $-0.36$ eV, which is comparable with related studies. 

By Bader charge analysis, we show that $A_3 B$ donates electrons during the H-adsorption; accordingly, an intermediate bonding between the covalent and ionic bonding forms. 
Interestingly, sites and materials with fewer electrons transferred to the H atom tend to obtain lower $\Delta G$. 
The difference in electronegativity explains such features and indicates that $A_3 B$ follows the $d$-band theory for transition metal-based catalysts. 
We can also explain the observations by projected band structures. 
The DOS spectrum of cases with lower $\Delta G$ features sharper H peaks, which indicate flat H bands, than dispersive bands across the energy spectrum. 
From a band theory perspective, we conclude that the localized H state indicates a weak interaction with the substrate, which is a good indicator according to the Sabatier principle. 

Finally, we conclude that fcc and hcp sites outperform the top site due to the existence of TSS and, consequently, stronger topological polarization. 
We discover that H interacts with TSS on the fcc and hcp site, and weak covalent bonding is formed, reflected by the flat H bands in the band structure. 
On the contrary, the charge transfer of the top site is happening between H and the bulk bands.
Once H interacts with the bulk bands, more hybridization between H and $A_3 B$ would occur, resulting in larger $\Delta G$. 
This work firstly suggests that the $A_3 B$ family are rare cases of semimetals to acquire $\pi$ Zak phase along the whole projected BZ; secondly, shows that Zak phase induced topological polarization is a promising platform to study electrocatalysts; thirdly, introduces a new viewpoint of the Sabatier principle; and finally unveils the mechanism of TSS toward HER in the $A_3 B$.

\section*{APPENDIX: EFFECT OF SPIN-ORBIT COUPLING}
\begin{table}[htbp]
\label{table:Gibbssoc}
\caption{The adsorption energy ($E_{\rm\text{ads}}$), Gibbs free energy difference ($\Delta G$) with spin-orbit coupling at room temperature of H on different adsorption sites (fcc, hcp, and top site) of the 10-layer $A_3 B$ (111) substrate. Also, the Bader charge analysis results of the top layer atoms are listed with charge differences before and after the adsorption given in the row below. Note that the number of charges for $A$ is the average between the three atoms on the surface.}
\begin{ruledtabular}
\begin{tabular}{c | c c | c c  |c c  }
substrate&  fcc& (eV)  &  hcp& (eV) & top& (eV)  \\
$A_3 B$   &  $E_{\rm\text{ads}}$& $\Delta G$& $E_{\rm\text{ads}}$& $\Delta G$  & $E_{\rm\text{ads}}$& $\Delta G$  \\
\hline
Pt$_3$Sn& -0.322& -0.064& -0.572& -0.348& 0.798& 1.024\\
Pd$_3$Sn& -0.355& -0.090& -0.667& -0.429& 0.993& 1.205\\
Ni$_3$Sn& -0.633& -0.358& -0.909& -0.655& 0.929& 1.137\\
Ni$_3$Ge& -0.542& -0.260& -0.972& -0.706& 0.826& 1.052\\
Ni$_3$Si& -0.494& -0.212& -0.987& -0.720& 0.630& 0.857\\
\hline
Pt      & -0.335& -0.116&       &       & -0.375& -0.115
\end{tabular}
\end{ruledtabular}
\end{table}


When considering the SOC, which is large in Pt-based systems, the topological phase of $A_3 B$ will shift from a $\pi$ Zak phase to a weak Dirac dispersion along the [111] direction \cite{Kim2017} due to the gap opening at $\Gamma$ and $R$ point (Fig. S6(a)).
In Fig. S6(b), we see the slab band structure of the constructed Wannier functions with SOC. Compared to Fig. \ref{fig:elec}(e), surface states remain in the gap with split Kramers degeneracy at all $\textbf{k}$ points except the time-reversal invariant point $\bar{M}$. Furthermore, DFT results are given in Fig.S6 (c) and, similar to spinless case (Fig. \ref{fig:elec}(d)), surface bands merge into bulk bands and surface states are hidden around the Fermi level. Interestingly, the band dispersion before and after the Pt$_3$Sn fcc-adsorption is not highly affected by the SOC (Fig. \ref{fig:elec}(f), \ref{fig:projected}(a), Fig. S6 (c), and Fig. S6 (d)). This indicates that the hidden surface states still interact with the H atoms and thus the SOC plays a negligible role. The existence of surface states guaranteed by the nontrivial $\pi$ Zak phase plays a significant role in the current scenario. Accordingly, in Table \ref{table:Gibbssoc}, we look into the effect of the SOC on the catalytic behavior by calculating $\Delta G$ with SOC included.  

First of all, we note that the calculated vibrational frequency with or without the SOC are close. Therefore, the changes in the resulting $\Delta G$ would mainly come from the changes in $E_{\rm\text{ads}}$. From Table \ref{table:Gibbssoc}, generally, the $\Delta G$ becomes smaller when SOC is considered.  Based on the compounds, the difference, due to the atomic number, is roughly 4.5\%, 0.9\%, 0.7\%, 0.4\%, 0.5\%, and 15\% for Pt$_3$Sn, Pd$_3$Sn, Ni$_3$Sn, Ni$_3$Ge, Ni$_3$Si, and Pt. In the previous paragraph, we argued the interaction between surface states and H atom is insignificantly affected by the SOC. Here, from Table \ref{table:Gibbssoc}, we can see that the SOC effect on catalytic behavior is roughly in the order of 10 meV, while the difference among adsorption sites is in the order of 100 meV (Table \ref{table:Gibbs}, \ref{table:Gibbssoc}). Thus, we conclude that the proposal regarding the role of TSS is still valid in the case of $A_3 B$ when considering the SOC. 

\section*{ACKNOWLEDGMENTS}
The authors thank Yuto Miyahara, Kohei Miyazaki, and Hiroshi Kageyama for fruitful discussions throughout this work. M.-C. J. and G.-Y. G. acknowledge the support from the Ministry of Science and Technology and the National Center for Theoretical Sciences (NCTS) of The R.O.C. M.-C. J. and G.-Y. G. are also grateful to the National Center for High-performance Computing (NCHC) for the computing time. M.-C. J. was supported by RIKEN's IPA Program. M. H. acknowledges financial support from JSPS KAKENHI Grants (No. 20K14390) and PRESTO, JST (JPMJPR21Q6). T. Y. acknowledges financial support from JSPS KAKENHI Grants (No. 22J23068). T. N. acknowledges financial support from PRESTO, JST (JPMJPR20L7). R. A. was supported by a Grant-in-Aid for Scientific Research (No. 19H05825) from the Ministry of Education, Culture, Sports, Science and Technology.


\bibliographystyle{apsrev}
\bibliography{TopoCat_abbr}
\end{document}